\documentclass[%
 reprint,
 superscriptaddress,
 amsmath,amssymb,floatfix,
aps,
]{revtex4-1}

\usepackage{sidecap}
\usepackage{ulem}
\usepackage{graphicx}
\usepackage{dcolumn}
\usepackage{bm}
\usepackage{textgreek}
\usepackage{floatrow}

\begin{document}


\title{Growth and Magnetotransport in Thin Film $\alpha$-Sn on CdTe}

\author{Owen Vail}
\author{Patrick Taylor}
\author{Patrick Folkes}
\author{Barbara Nichols}
\affiliation{US Army Research Laboratory, Adelphi, MD 20783, USA}
\author{Brian Haidet}
\author{Kunal Mukherjee}
\affiliation{Materials Department, University of California Santa Barbara, Santa Barbara, California 93106, USA\\}
\author{George de Coster}
\affiliation{US Army Research Laboratory, Adelphi, MD 20783, USA}

\date{\today}

\begin{abstract}
We report growth and characterization of epitaxial $\alpha$-Sn thin films grown on CdTe(111)B. Noninvasive techniques verify the film's pseduomorphic growth before fabrication of magnetotransport devices, overcoming ex-situ obstacles on uncapped films for measurement in the Hall bar geometry. We identify a transition to metallic behavior at low temperature with large magnetoresistance, high mobility, and quantum oscillations tentatively suggesting an n-type Dirac semimetallic channel. A parallel p-type dopant channel with high carrier density is seen to dominate at thinner film thicknesses.  Careful preparation of the CdTe surface before growth is considered crucial to attain a low dopant density and accessible topological states on an insulating substrate.
\end{abstract}

\pacs{Valid PACS appear here}
\maketitle


\section{\label{sec:intro}Introduction}

The diamond-cubic phase of tin (Sn), known as $\alpha$-Sn, has long been of experimental and theoretical interest due to its inverted electronic bulk band structure \cite{Groves1963,Kufner2015}. Such an inversion in a elemental material makes $\alpha$-Sn singularly exciting to investigate as a topological material because the strong spin orbit coupling required to cause a topological band inversion generally limits the choice of materials for study to complex compounds or heterostructures such as an engineered quantum well, ternary or quaternary compounds, or use of poisonous elements like mercury \cite{Barfuss2013,Xu2017,Hasan2010,franz2013topological,Bernevig2006,Fu2007,Fu2011}. However, spin orbit coupling strength can be increased even in single element crystals by picking heavier constituent elements. As an example, graphene is effectively a single element 2D Dirac semimetal (DSM) due to its low spin orbit coupling \cite{KaneMele2005}, while stanene is a two-dimensional topological insulator with a large inverted band gap and well defined edge states \cite{Xu2013,Xu2018}.

\begin{figure}[!b]%
\centering
  \includegraphics*[width=0.77\textwidth]{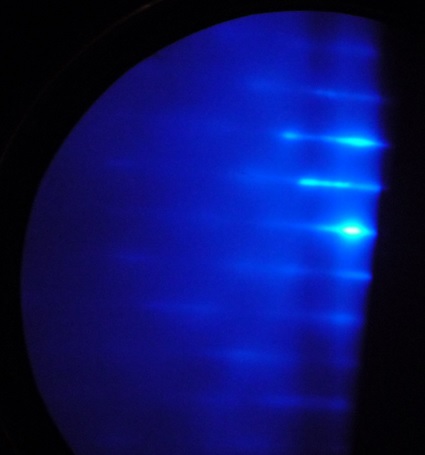}%
  \caption{RHEED pattern observed during the MBE $\alpha$-Sn growth of $\alpha$-Sn along a $\left<110\right>$ azimuth provides evidence for coherent epitaxial alignment with CdTe(111)B and smooth surface morphology.}
    \label{fig1}
\end{figure}

\begin{figure*}[!t]%
\floatbox[{\capbeside\thisfloatsetup{capbesideposition={right,bottom},capbesidewidth=0.18\textwidth}}]{figure}[\FBwidth]
{\caption{Raman spectroscopy and optical inset of $\alpha$-Sn thin film.  The $1\Gamma$ peak in intensity at 198 cm$^{-1}$ is characteristic of the diamond cubic phase of Sn. Combined with the uniform morphology apparent optically (inset), this indicates a high quality film growth. An additional peak near 379 cm$^{-1}$ and background between 200 and 350 cm$^{-1}$ are consistent with the two phonon overtones in $\alpha$-Sn(111) \cite{Iliev1977}.}}
{\includegraphics*[width=.77\textwidth]{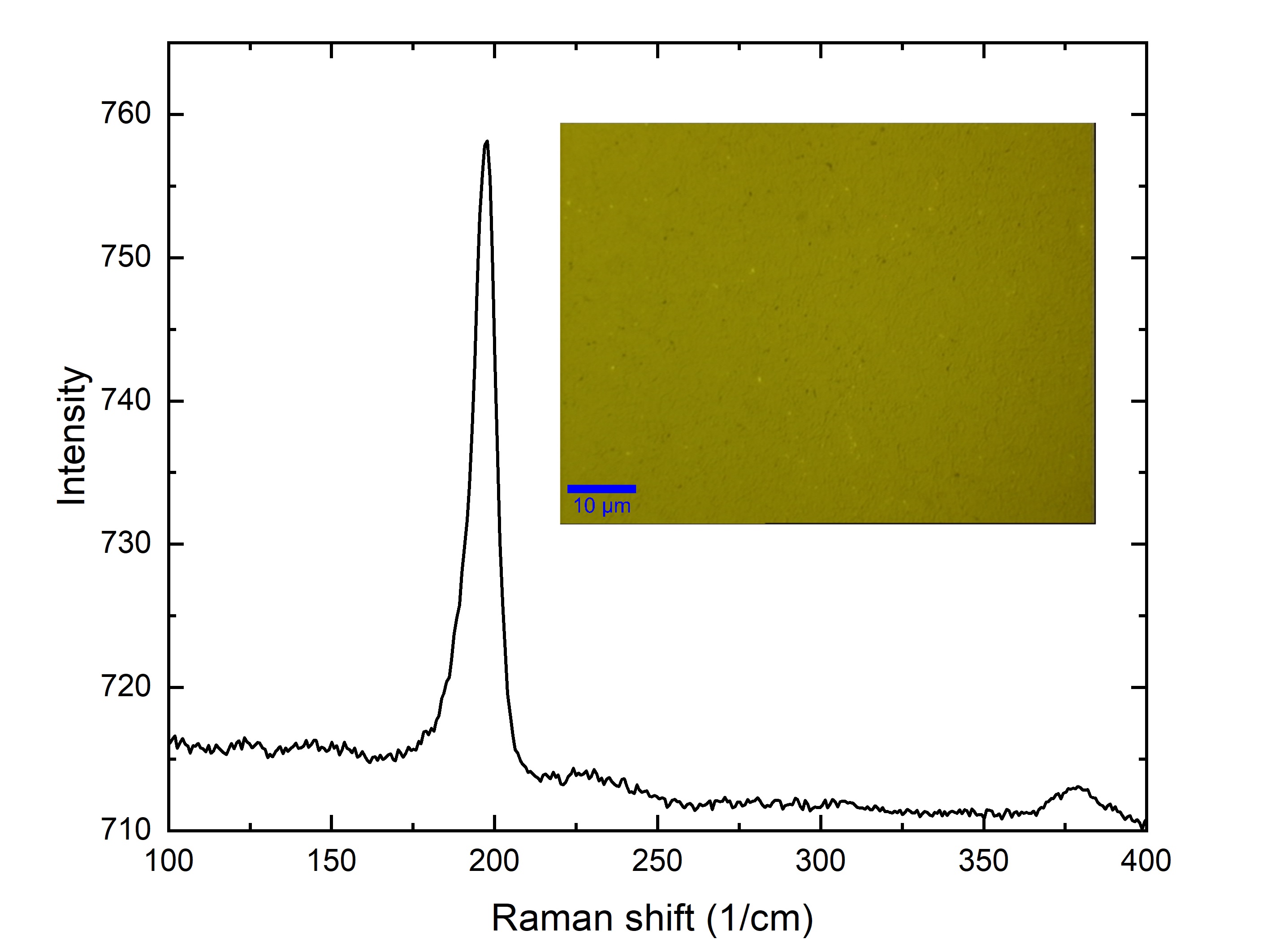}}
    \label{fig2}
\end{figure*}

Replacing carbon atoms in a three dimensional diamond cubic lattice with Sn can also achieve non-trivial topological physics. Indeed, through different strain, orientation, and thickness configurations one can achieve two and three dimensional topological insulator and DSM states in $\alpha$-Sn thin films \cite{decoster2018,Huang2017,Abrikosov1971}. The ability to produce and process high quality Sn films will be a significant boon in achieving the diverse applications of topological physics such as efficient nonvolatile memory and low voltage devices \cite{LiSemenov2014,Semenov2012,Semenov2014,Mellnik2014}. The persistence of the topological states near room temperature could enable these applications to be employed in all manner of environments \cite{Rojas2016,Li2014,Ohtsubo2013}.

The diamond cubic $\alpha$-Sn phase is generally stable below 13.2$^\circ$C, otherwise it undergoes a structural transition to the $\beta$-Sn phase. The critical temperature of this phase transition can be pushed above room temperature when the $\alpha$-Sn is grown in thin films by molecular beam epitaxy (MBE) due to the stabilizing effect of a diamond cubic substrate \cite{Farrow1981}. Unlike the metallic $\beta$-Sn phase, unstrained bulk $\alpha$-Sn is a zero-gap semiconductor. Tensile epitaxial strain and thickness dependent quantum confinement can create band gaps in thin films that are large enough to limit thermal excitations even at room temperature \cite{Rogalev2017,Barfuss2013}, while compressive strain can cause sufficiently thick films to be DSMs \cite{Xu2017}. The various topological phases achievable in Sn structures critically depend on the spin orbit coupling and mass Darwin effect induced band inversion between the $\Gamma_8^+$ and $\Gamma_7^-$ bands, revealing nontrivial topology in transport measurements when the Fermi level is positioned appropriately \cite{Kufner2015}. We remark that $\alpha$-Sn can be thought of as the inversion symmetric case of HgTe, which hosts many of the same topological states predicted in $\alpha$-Sn. However, unlike $\alpha$-Sn, compressive epitaxial strain does not yield a DSM state in HgTe due to its broken inversion symmetry \cite{Zaheer2013}.

An insulating substrate is of critical importance to the electronic application of thin film topological materials. The zincblende crystal structure of CdTe has a lattice constant ($a = 6.481$ \AA) well matched to $\alpha$-Sn ($a = 6.4892$ \AA), and is insulating with a 1.6 eV band gap \cite{Adachi_book,Novik2005}. The (111) projection of the diamond-cubic lattice is particularly exciting because it is honeycomb, such that pseudomorphic monolayer growth on this surface may yield stanene. By epitaxially growing $\alpha$-Sn on CdTe(111) and tuning the film thickness, it may be possible to observe a variety of quantum confinement induced topological states, including monolayer stanene. Fundamental growth and fabrication techniques are lacking for the $\alpha$-Sn/CdTe material system when compared with its InSb counterpart. However, the narrow band gap of InSb imposes parallel transport channels to the $\alpha$-Sn making it ill-suited for device engineering. Transport devices fabricated from $\alpha$-Sn/CdTe heterostructures give insight into its promise as a practical electronic platform.

\section{\label{sec:method}Experiment}

Substrates of Te terminated CdTe(111)B surfaces were selected for $\alpha$-Sn growth, as the Te atoms bond covalently with Sn, enforcing the diamond lattice formation of $\alpha$-Sn formation. Growth of $\alpha$-Sn on the Cd terminated CdTe(111)A is more difficult due to the metallic bonding between Sn and Cd atoms. The CdTe surface is treated with a bromine etch to attain a Te-rich layer that will thermally desorb during annealing. The substrate is annealed in 10\textsuperscript{-10} Torr at 250-300 $^{\circ}$C, and Reflection High-Energy Electron Diffraction (RHEED) oscillations are taken to characterize the pristine (111)B surface. In order to promote the formation of an abrupt interface, the substrate is allowed to cool in vacuum to a temperature of -15 $^{\circ}$C before $\alpha$-Sn is grown through the beaming of Sn molecules onto the cold hexagonal surface. RHEED oscillations taken during MBE growth indicate a well-ordered crystal formation along a $\left<110\right>$ azimuth (Fig. \ref{fig1}). A strong RHEED pattern indicates coherent epitaxial alignment with the CdTe(111)B substrate and correlates with smooth surface morphology.

After the films are removed from vacuum, Raman spectroscopy is performed using a Witech alpha300A microscope and a 532 nm wavelength laser operating at room temperature with 1 mW beam power in a 1 $\mu$m spot size and a 10 s measurement time to avoid heating of the sample.  We identify the $\alpha$-Sn 1$\Gamma$ peak at 198 cm\textsuperscript{-1} (Fig. \ref{fig2}) \cite{Buchenauer1971} and observe the surface uniformity in order to confirm high quality $\alpha$-Sn growth. The $\alpha$-Sn Raman peak is still prevalent after months of exposure without a capping layer, which bodes well for ex-situ processing and realistic applications that cannot be done under vacuum.  By increasing the beam intensity over 50 mW, we are able to cause local transitions to $\beta$-Sn optically visible as patches of white on a grey Sn surface with no Raman peak near 198 cm\textsuperscript{-1}.

\begin{figure}%
    \centering
    \includegraphics*[width=1.1\textwidth]{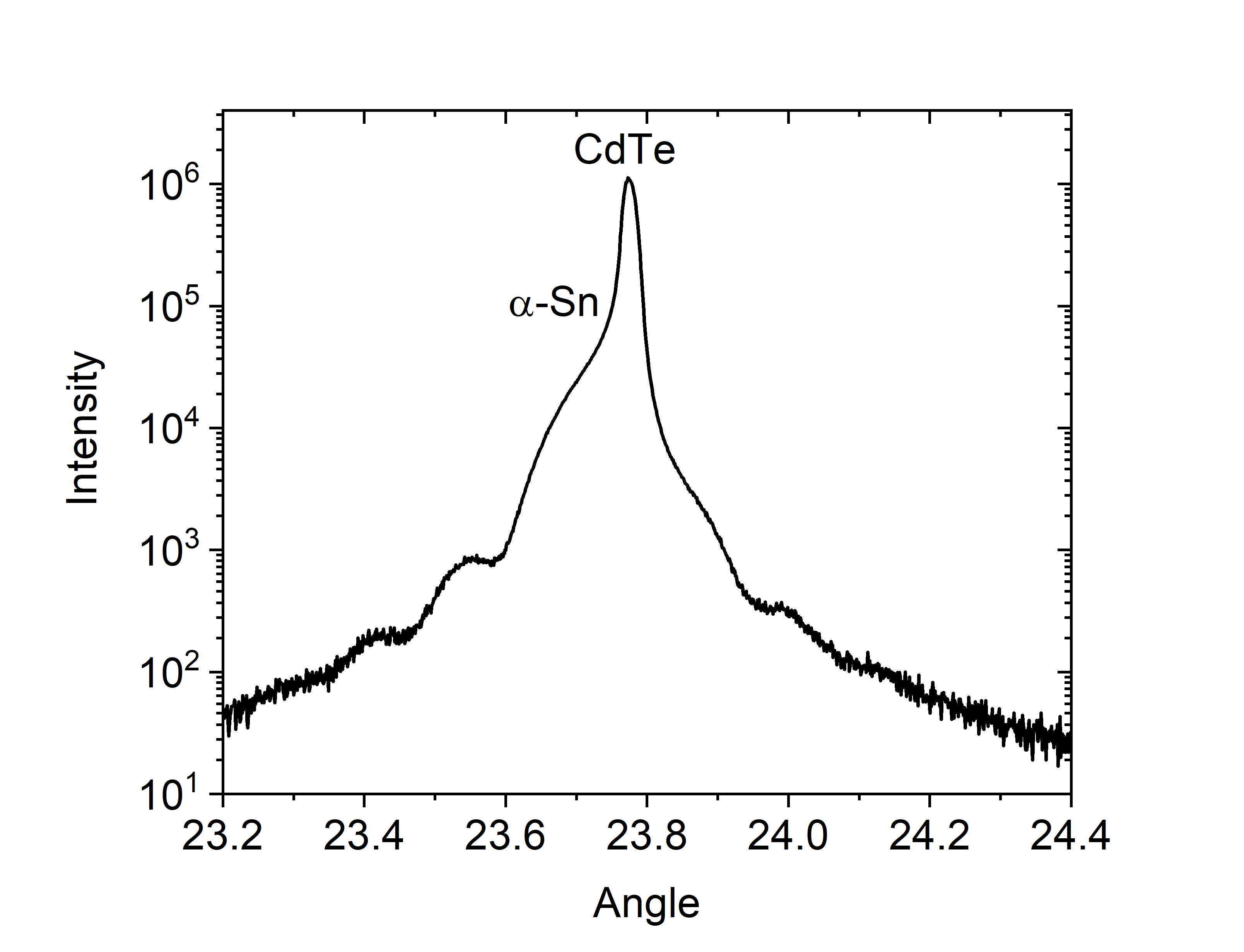}
    \caption{$\theta$-$2\theta$ XRD spectra of $\alpha$-Sn grown on a CdTe(111) substrate.  The minimal shift between $\alpha$-Sn and CdTe peaks implies pseudomorphic growth with compressive strain. Kiessig fringes appear symmetrically around the $\alpha$-Sn peak.}
    \label{fig3}
\end{figure}

While shifts in the Raman spectrum give some indication of epitaxially compressive strain in the film \cite{Bell1972}, X-ray diffraction (XRD) provides more detailed information with respect to the $\alpha$-Sn diamond-cubic crystal structure on the CdTe substrate. Fig \ref{fig3} shows the symmetric $\theta$-$2\theta$ (111) X-ray reflection from this sample. The main peak is a convolution of signals from the CdTe(111) and $\alpha$-Sn(111), with the CdTe being responsible for the mammoth peak and the $\alpha$-Sn arising as a shoulder on the left-hand side. The position of these peaks implies pseudomorphic crystal growth with a lattice constant for the $\alpha$-Sn film which exceeds that of the CdTe substrate. If we assume the unit cell volume is conserved, the lattice distortion is consistent with compressive strain in the film causing Poisson expansion in the growth direction. Kiessig fringes appear as symmetric ripples about the $\alpha$-Sn peak \cite{Kiessig1930}. These fringes give information about the thickness of the layer, which we approximate as 45 nm. Though this is not an exact measurement of thickness, it is in agreement with the estimated 48 nm thickness extrapolated from growth rates. According to convention, a bilayer of $\alpha$-Sn(111) is defined as a single layer of stanene, which is a buckled honeycomb lattice of Sn atoms. The distance between bilayers is 3.748 \AA, implying a thin film thickness of approximately  120 bilayers.

Thin films of $\alpha$-Sn have been shown to transition to $\beta$-Sn at temperatures near 70$^{\circ}$C \cite{Menendez1984}. With this in mind, a room temperature vacuum anneal is used to cure polymer for lithographic device design. Dry etching of Hall bars using argon plasma affords precision and abrupt mesa sidewalls. The resulting mesoscopic (20$\times$10 $\mu$m$^2$) Hall bars are contacted using electron beam evaporation. As the CdTe substrate is particularly brittle, a metallic adhesion layer and ball bonding technique helps guarantee the success of subsequent device bonding. The resulting packaged devices resemble end-user electronics that can be plugged in like a computer chip.

\begin{figure}%
    \centering
    \includegraphics*[width=1\textwidth]{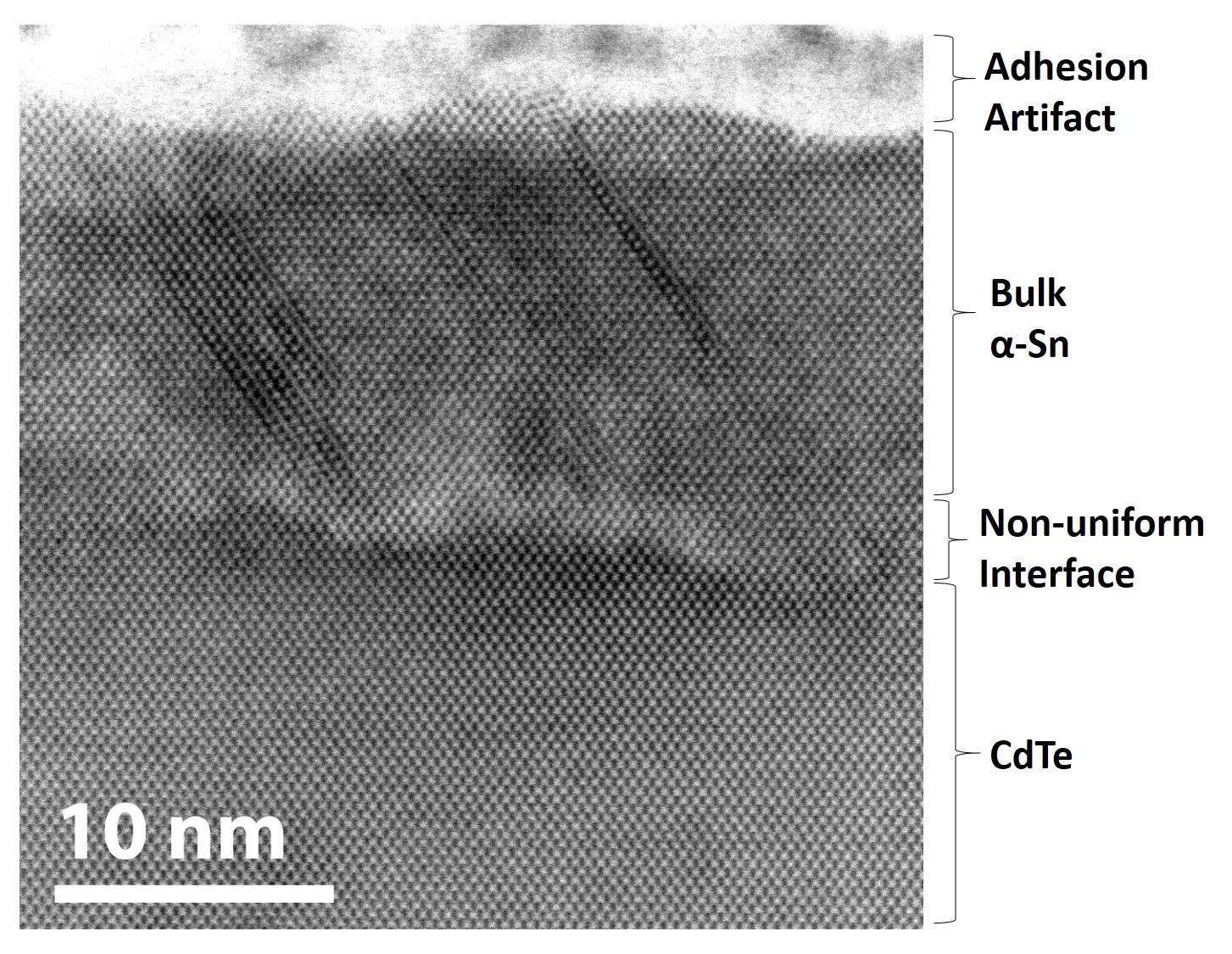}
    \caption{Transmission electron microscopy image of $\alpha$-Sn/CdTe(001) heterostructure. Irregularities of the substrate surface critically affect the dopant density of $\alpha$-Sn in thin film samples. Stacking fault defects appear in the bulk $\alpha$-Sn. Coloration near the interface is indicative of non-uniform strain across the epitaxial film.}
    \label{fig4}
\end{figure}

Achieving an abrupt $\alpha$-Sn/CdTe interface can be particularly difficult \cite{Zimmermann1997} especially when compared with a substrate like InSb \cite{Xu2017}. Furthermore, the irregularity of the surface quality of industrially purchased CdTe material greatly affects the reproducibility of high-quality $\alpha$-Sn films. Transmission electron microscopy images of $\alpha$-Sn/CdTe interfaces reveal non-uniform strain and defects (Fig. \ref{fig4}) as a result of imperfect surface preparation. However, the low carrier density of the semi-insulating substrate (less than 10$^{12}$ cm$^{-3}$ with 10$^{14}$ cm$^{-3}$ dopant density) allows for a simplified model of the conduction pathways in the system by assuming that the contribution of the substrate to the transport properties is negligible. With this in mind, Hall bars fabricated on samples of $\alpha$-Sn on CdTe are studied in a liquid helium cryostat under application of a perpendicular magnetic field.

We analyze the transport properties of a particular $\alpha$-Sn/CdTe(111) sample. While the $\alpha$-Sn thickness extrapolated from the growth rate for this sample is 48 nm, a layer of stannous oxide is expected to form from the topmost layers of Sn atoms due to atmospheric oxidation, resulting in a film thickness of approximately 46 nm \cite{Salt1956}, which we will use in subsequent calculations. This is far thicker than the thin films in recent transport studies of topologically insulating $\alpha$-Sn \cite{Barbedienne2018} and we should expect a 46 nm thick $\alpha$-Sn film grown on CdTe(111)B to be a DSM according to theoretical calculations and high resolution electron energy loss spectroscopy measurements of thin film $\alpha$-Sn band structures \cite{decoster2018,Huang2017,Takatani1985}. On the other hand, this thickness approaches an intermediate value proposed by Tu et. al. for the dominance of dopants produced from outdiffusion of substrate elements \cite{Tu1989Grow}. The prominence of these dopants depends critically on the CdTe surface preparation \cite{Yi1998}.

\section{\label{sec:exp}Results and Discussion}

\subsection{\label{sec:temp}Temperature Dependence of the Longitudinal Resistance}

\begin{figure}[!t]%
  \centering
  \includegraphics*[width=1.1\textwidth]{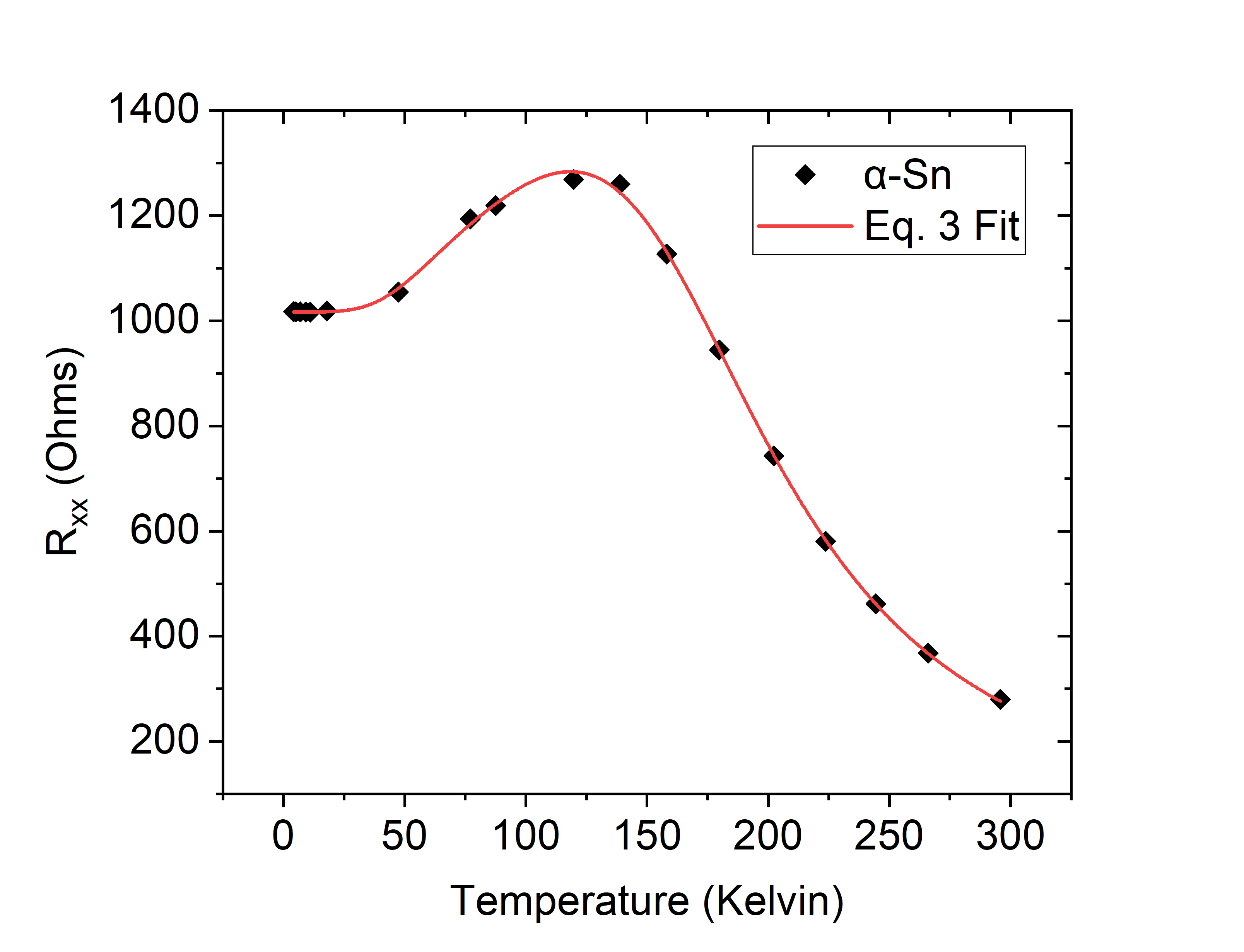}%
  \caption{Temperature dependence of $\alpha$-Sn device resistance on CdTe(111)B. The device exhibits semiconducting behavior near room temperature and metallic behavior at lower temperatures. There is good agreement between the full curve and Eq. \eqref{RT}, with a thermal activation energy of 81 meV.}
    \label{fig5}
\end{figure}

By applying 10 $\mu$A of direct current and measuring the longitudinal resistance of the Hall bar using a 4-point setup, the linear current-voltage curve is found to pass through the origin with $\pm$ 0.1 $\Omega$ of accuracy and stability. As the temperature of the device is reduced, the resistance initially increases exponentially with temperature, similar to the behavior of a semiconductor \cite{AshcroftMermin}. Using a combination of liquid nitrogen and liquid helium measurements, we establish a full temperature range with two distinct regimes, one with metallic behavior and one semiconducting (Fig. \ref{fig5}). We use an Arrhenius model for the semiconducting resistance $R_{\text{semi}}$ near room temperature with thermal carriers excited across a gap:

\begin{equation}
R_{\text{semi}}(T) = R_\infty e^{\frac{E_g}{k_B T}}~,
\label{Rsemieqn}
\end{equation}
where $k_B$ is the Boltzmann constant, $R_\infty$ is the resistance at high temperature, and $E_g$ is the thermal activation energy.

We can also understand the metallic channel resistance $R_{\text{metal}}(T)$ through the Bloch-Gr\"{u}neisen model for the resistance of electrons interacting with phonons in a metal \cite{ZimanBook}:
\begin{equation}
\frac{1}{R_{\text{metal}}(T) - R_0} = \frac{1}{C_1 T^5} + \frac{1}{C_2 T} ~.
\label{Rmeteqn}
\end{equation}
Here $C_1$ and $C_2$ are related to the prevalence of electron-phonon interactions in the metallic channel and the geometry of the Hall bar, and $R_0$ is the resistance at $T=0$ due to electron-defect scattering.  The $T^{5}$ scaling of the first term in Eq. \eqref{Rmeteqn} corresponds to the low temperature resistivity of a three dimensional metal (or a doped DSM \cite{DasSarma2015,Rodionov2015,Bloch1930}), while a two dimensional metal would have a $T^4$ scaling \cite{Hwang2008}. As the best fit uses a $T^5$ scaling and minimizes any $T^4$ component, we conclude that the dominant transport channel is three dimensional. This is consistent with the topological surface state of DSM $\alpha$-Sn being buried in the bulk valence and sub-valence bands \cite{Rogalev2017}.

If we assume a metallic channel in parallel with the semiconducting one, the total resistance as a function of temperature is given by
\begin{equation}
\dfrac{1}{R(T)} =\dfrac{1}{R_\text{semi}(T)} + \frac{1}{R_{\text{metal}}(T)} ~,
\label{RT}
\end{equation}

By fitting the full temperature range with Eq. \eqref{RT}, we find $C_1=2.84\times10^{-7}$ \textOmega /K$^5$, $C_2=2.77$ \textOmega /K, $R_0=1017$ \textOmega, and $R_\infty=13.6$ \textOmega, and extract a $E_g = 81$ meV thermal excitation energy. This activation gap obtained for the semiconducting region falls within accepted values of the L$^+_{6c}$-$\Gamma^+_{8vc}$ indirect band gap of unstrained $\alpha$-Sn \cite{Brudevoll1993}, suggesting that indirect interband transitions are responsible for the semiconducting behavior. Previous studies on $\alpha$-Sn found a significant temperature dependence to this gap \cite{Hoffman1989} but attempts to include this parameter in Eq. \eqref{Rsemieqn} do not affect the other parameters and the magnitude of temperature dependence is minimized, resembling the results found by Lavine and Ewald \cite{Lavine_1971} at low temperature. The best fit reveals a reduction in $E_g$ of only 5 meV at room temperature. We can expect the epitaxial strain to play a significant role in the indirect band gap \cite{Hoffman1989}, especially at these thin film thicknesses, and may be a potential source of deviation from previous reports. Thinner film thicknesses showed an increased gap energy, consistent with quantum confinement. In summation, we find the behavior of $R(T)$ in Eq. \eqref{RT} to be dominated by $R_\text{semi}(T)$ near room temperature and $R_\text{metal}(T)$ near and below 70K, implying that studies of the metallic behavior may be accessible at liquid nitrogen temperatures.  We should expect the low temperature transport properties of thin film DSMs to be metallic in nature, dominated by intraband scattering mechanisms \cite{DasSarma2015,Rodionov2015}. In fact, the high temperature crossover to semiconductor-like transport is consistent with the R-T curves measured in other thin film DSMs such as Na$_3$Bi and Cd$_3$As$_2$ \cite{Liu2017Na3Bi,Zhou2016,Li2015Cd3As2,Li2016Cd3As2}.

\begin{figure}%
\centering
  \includegraphics*[width=0.9\textwidth]{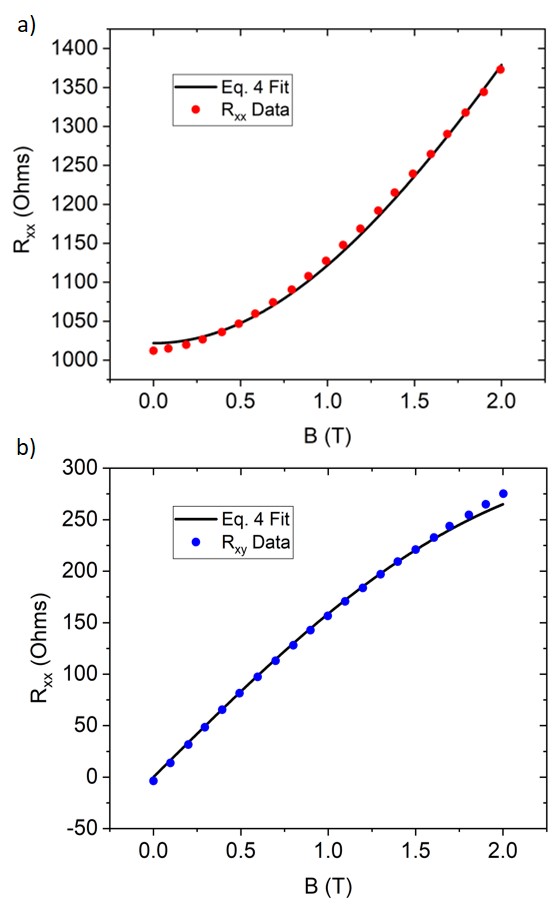}%
  \caption{Magnetic field dependence of the longitudinal and transverse resistance at 4.2 K for an $\alpha$-Sn thin film grown on CdTe(111)B.  The nonlinear behavior of a) R$_{xx}$ and b) R$_{xy}$ is expected in real metals and semimetals, and indicates multiple carrier channels. The curves are fit with Eq. \eqref{res} in order to extract the carrier densities and mobilities.}
    \label{fig6}
\end{figure}

\begin{figure}%
\centering
  \includegraphics*[width=1.05\textwidth]{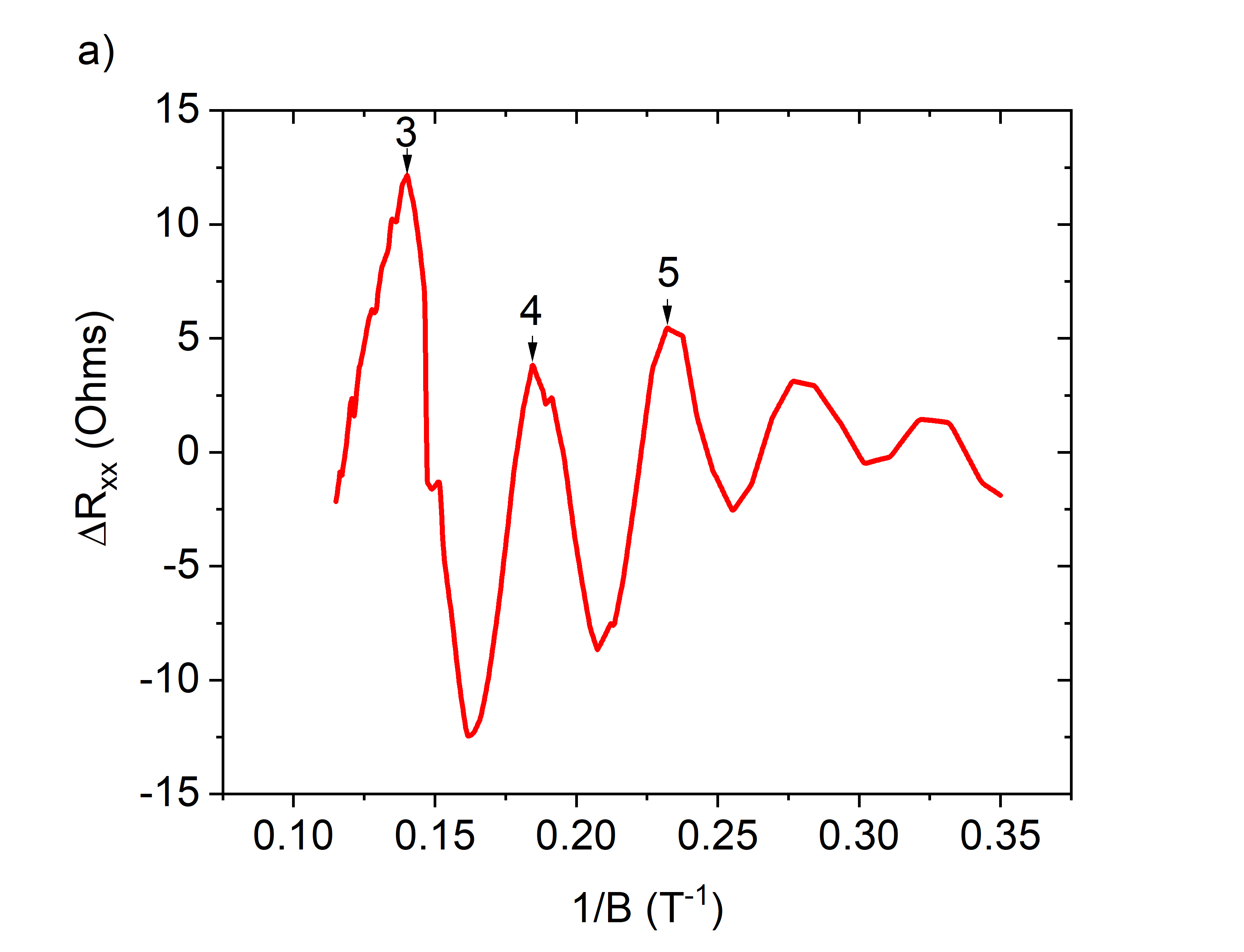}%
  \\
  \includegraphics*[width=1.05\textwidth]{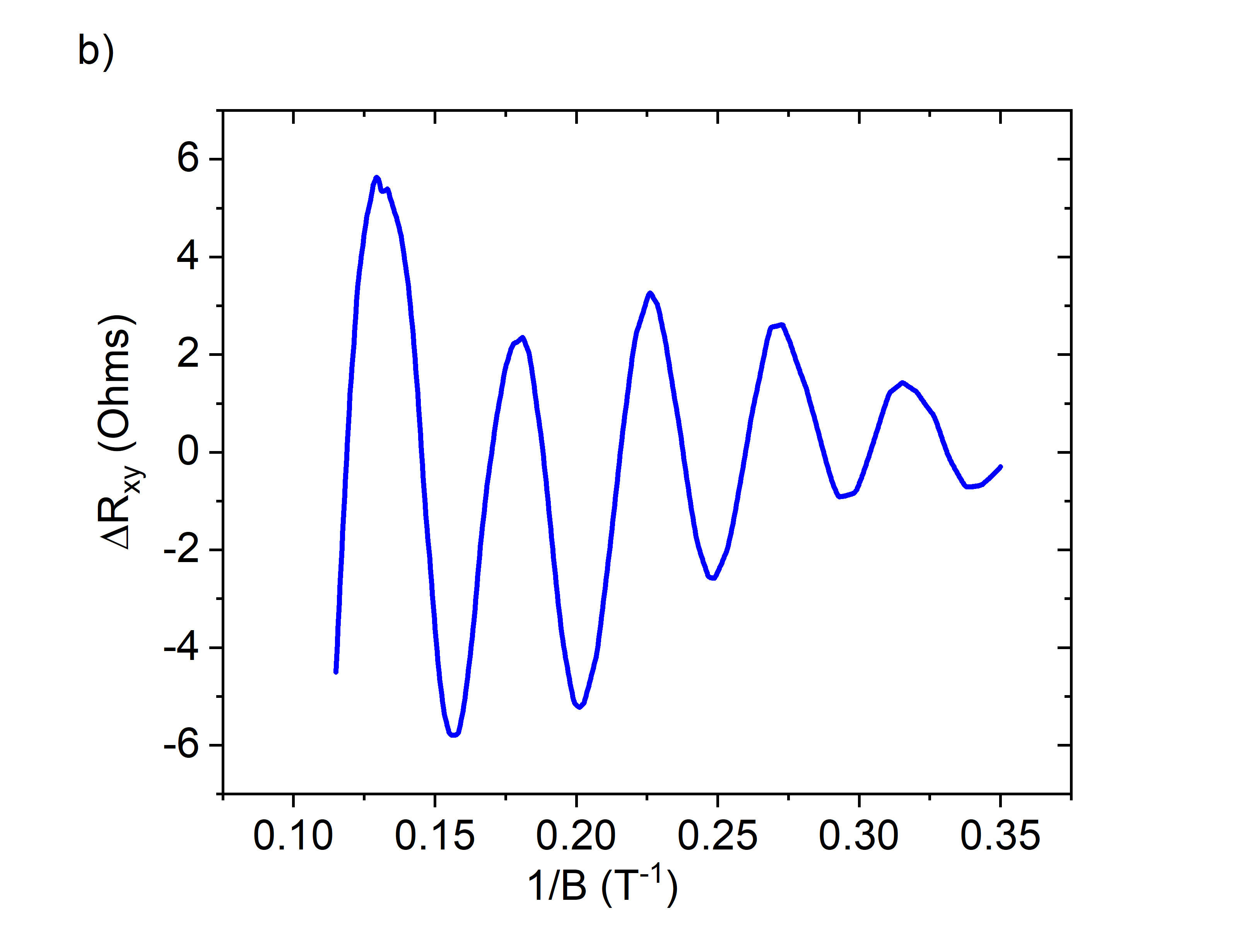}%
  \\
  \includegraphics*[width=1.05\textwidth]{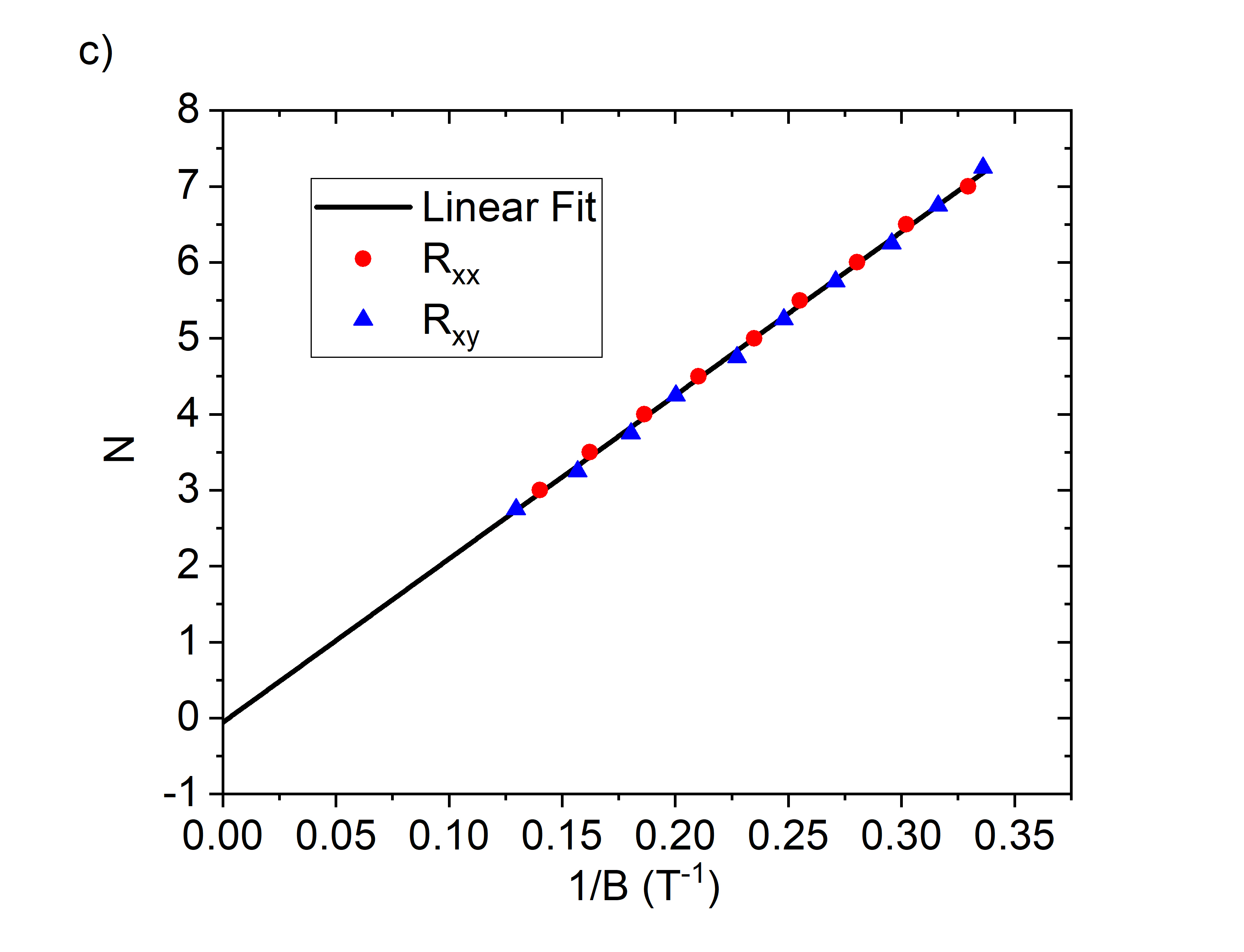}%
  \caption{High magnetic field Shubinkov de-Haas oscillations of a) $\Delta R_{xx}$ and b) $\Delta R_{xy}$. Extrema are assigned to their corresponding fractional LLs and used to construct a linear fit c) with a slope of 21.56 T and an intercept of $-0.063$. This intercept provides direct evidence of Dirac dispersion via Eq. \eqref{sdh}.}
    \label{fig7}
\end{figure}

\subsection{\label{sec:mag}Magnetic Field Dependence of the Longitudinal and Hall Resistances}

At low temperature, the device reached a finite resistance that no longer depended on temperature. Upon application of a low magnetic field, the Hall resistance ($R_{xy}$) increased, consistent with an electron-like charge carrier. The longitudinal resistance ($R_{xx}$) exhibited a quadratic dependence, which is common for materials with multiple carriers \cite{AshcroftMermin}. The high field data reveals unsaturating magnetoresistance on the order of $200\%$ at 9 T, which has been seen to occur in approximately electron-hole compensated systems and topological semimetals more generally \cite{Hu2019}. As such, we restrict our two-carrier model to a simultaneous fit of the low-field $R_{xx}$ and $R_{xy}$ data up to 2 T where quantum effects are not significant. The longitudinal and transverse resistivities are found by inverting the two-carrier conductivity tensor:

\begin{eqnarray}
\sigma_{xx}=\frac{n_ne\mu_n}{1+(\mu_nB)^2}+\frac{n_pe\mu_p}{1+(\mu_pB)^2}~,\notag\\
\sigma_{xy}=\frac{n_ne\mu_n^2B}{1+(\mu_nB)^2}-\frac{n_pe\mu_p^2B}{1+(\mu_pB)^2}~,\notag\\
\rho_{xx}=\frac{\sigma_{xx}}{\sigma_{xx}^2+\sigma_{xy}^2}~, \qquad \rho_{xy}=\frac{\sigma_{xy}}{\sigma_{xx}^2+\sigma_{xy}^2}~.
\label{res}
\end{eqnarray}

Here $e$ is the electron charge, $B$ is the applied field, $n_n$ and $\mu_n$ are the n-type carrier density and mobility, while $n_p$ and $\mu_p$ are the p-type carrier density and mobility. Three dimensional resistances are found from these quantities using the geometric prefactors. The resulting fit (Fig. \ref{fig6}) gives a dominant n-type channel with $n_n=2.68\times10^{17}~\text{cm}^{-3}$  and $\mu_n=5900~\text{cm}^2/\text{Vs}$, as well as a p-type dopant channel with $n_p=1.83\times10^{18}~\text{cm}^{-3}$  and $\mu_p=600~\text{cm}^2/\text{Vs}$. The value of n-type carrier density and mobility is consistent with earlier studies of $\alpha$-Sn crystals \cite{Lavine_1971,Broerman1970}.

Under a strong magnetic field ($>2$ T) Shubnikov de-Haas (SdH) oscillations were observed. Background polynomials are subtracted from $R_{xx}$ and $R_{xy}$, giving $\Delta R_{xx}$ and $\Delta R_{xy}$, such that the periodic oscillations can be clearly seen as a function of $1/B$ (Fig \ref{fig7}). Confirming peak-to-peak measurements with a Fourier transform reveals an oscillation period of $F = 21.56$ T. If we assume an isotropic Fermi surface in a three dimensional semimetal at low temperature, we can calculate a SdH carrier density $n_{SdH}$ using:
\begin{equation}
n_{SdH}=2\frac{4\pi}{3}\frac{k_F^3}{(2\pi)^3}=\frac{1}{3\pi^2}\left(\frac{2eF}{\hbar}\right)^{3/2}~,
\label{sdh}
\end{equation}
where $E_F$ is the Fermi energy, $\omega_c$ is the cyclotron frequency, and $\hbar$ is the reduced plank constant. The resulting carrier density is $n_{SdH}=5.66\times10^{17}$ $\text{cm}^{-3}$. While this carrier density is not in precise agreement with the Hall analysis, a more realistic calculation of the carrier density factoring in the anisotropy of the Fermi surface in the $k_z$ direction would be expected to improve the agreement between the two analyses \cite{Huang2017}.

SdH oscillations are a consequence of Landau quantization of low carrier density states. Sequential emptying of Landau levels (LLs) leads to oscillations in the conductivity, such that minimas in $\sigma_{xx}$ correspond to integer LLs. By noting $\rho_{xx}\gg\rho_{xy}$ at all fields, this implies that $\sigma_{xx}\gg\sigma_{xy}$ and $\rho_{xx}\sim 1/\sigma_{xx}$ by Eq. \eqref{res}. Therefore the maxima of $\Delta R_{xx}$ are assigned to integer LL indices (N), while minima are assigned to half integers \cite{Xiong2012}. Extrema of $\Delta R_{xy}$ are phase-shifted by 1/4 \cite{Barbedienne2018,Xiong2012}, and thereby correspond to quarter integer Landau levels (N+1/4 for the minima and N+3/4 for the maxima of the transverse resistance). The Lifshitz-Kosevich theory describes the quantum oscillations through a cosine such that \cite{Xiao2015,Wright2013,Huang2017,Shoenberg1984}
\begin{equation}
\Delta\rho_{SdH} = A(T,B)\cos\left[2\pi\left(\frac{F}{B} - \frac{1}{2} + \frac{\Phi_B}{2\pi} - \delta\right)\right]~,\\
\label{SdH}
\end{equation}
where $\delta$ is a phase shift determined by the dimensionality, with $\delta=1/8$ corresponding a 3D corrugated Fermi surface and $\delta=0$ corresponding to a quasi-2D cylindrical Fermi surface \cite{Murakawa2013}. A(T,B) is an amplitude of oscillation while $\phi_B$ is the Berry phase related to the type of dispersion in the material. While the Land\'{e} g-factor is expected to be large for heavy elements like Sn \cite{Song1990}, we can expect A(T,B) to be positive due to the small effective mass of the conduction band \cite{Groves1970}. Therefore, we take the maxima to be integer LLs and fit the index with

\begin{equation}
N = \frac{F}{B} - \frac{1}{2} + \frac{\Phi_B}{2\pi} - \delta~.\\
\label{Berry}
\end{equation}

We find an intercept of -0.063 such that the Berry phase is close to $\pi$ with $\delta$ between 0 and $1/8$, suggesting a quasi-3D Dirac semimetal.  To characterize the Berry phase and the Dirac nature of the film, temperature and angular dependence of the SdH oscillations would provide an exploration of the Fermi surface while measurement of other exotic effects such as an anomalous Hall effect or those resulting from a chiral anomaly would give further indications of the nontrivial nature of the material.  In particular, resolving the sign of A(T,B) would unambiguously determine the phase offset in the quantum oscillations. However, even then a $\pi$ offset of this nature should not be taken as smoking gun evidence for a DSM \cite{Alexandradinata2018}.

The quantum confinement regime in $\alpha$-Sn grown on CdTe required for gapping out the bulk and producing a quasi-topological insulator state should be present below 40 nm \cite{decoster2018}, in contrast to growth on InSb which has only been shown to produce 3D topologically insulating $\alpha$-Sn below 10 nm \cite{Barbedienne2018}. However, when measuring 15 nm thick $\alpha$-Sn grown on CdTe, we find a single dominant channel with p-type carrier density of $6\times10^{19}~\text{cm}^{-3}$  and a mobility of $100~\text{cm}^2/\text{Vs}$, indicating that dopants dominate transport at this thickness. This is consistent with previous reports that thin film growth of $\alpha$-Sn on CdTe depends critically on substrate preparation \cite{Tu1989Grow, Yi1998}. With a dutifully prepared CdTe growth surface, electrically isolated elemental topological insulator $\alpha$-Sn stable in atmosphere without a capping layer may be accessible. 

The CdTe strain induced overlap in the $\Gamma_8^+$ bands of $\alpha$-Sn has been predicted to make it a DSM at these thicknesses \cite{Huang2017}.  However, previous studies of SdH oscillations in $\alpha$-Sn grown on CdTe did not identify DSM behavior and their SdH analysis did not discuss the intercept of Eq.\eqref{Berry}, instead assuming a parabolic band structure \cite{Tu1989SdH, Song1990}. Reference \cite{Song1990} also identified the oscillation maxima with integer LLs and appears to have plotted a non-trivial intercept, but assumed a trivial Berry phase factor in their analysis. As these previous reports are inconclusive on this point, the present results thereby suggest DSM behavior in this heterostructure for the first time using magnetotransport methods, however more extensive research is needed to confirm this.  

\section{\label{sec:conc}Conclusion}

In conclusion, we epitaxially grew and characterized thin film $\alpha$-Sn on CdTe(111)B. Non-invasive techniques were used to verify the high quality pseudomorphic growth of the film, followed by direct transport measurement. Films were found to have semiconducting behavior at room temperature and metallic behavior at low temperature with both n- and p-like carriers. The phase offset of the SdH quantum oscillations was characterized by extrapolating to the zero-energy Landau Level, and the result suggests a dominant DSM channel. By exploring the challenges associated with growth and fabrication of single element $\alpha$-Sn thin film devices on insulating CdTe, we provide a potential platform for future topological electronics based on this heterostructure. Moreover, the results presented indicate that capping and single digit Kelvin temperatures may be unnecessary to access the DSM regime of $\alpha$-Sn, opening the doors to a host of practical device applications. By refining the CdTe surface preparation method, we hope to introduce quantum confinement to access a wide array of topological device engineering. As such, further growth and transport studies on samples of this nature will help identify the best route toward a single-element topological system that is functional in the field.

\section*{\label{sec:ack}Acknowledgments}
This research was partially supported by a Laboratory University Collaborative Initiative award provided by the Basic Research Office in the Office of the Under Secretary of Defense for Research and Engineering.

\bibliographystyle{aprev}
\bibliography{aSn}

\end{document}